\documentclass[10pt]{article}
\usepackage{amssymb}
\usepackage{amsthm}
\usepackage{authblk}
\usepackage{setspace}
\usepackage[textwidth=6.5in,textheight=8.25in,centering]{geometry}
\usepackage[pdftex]{graphicx}

\onehalfspace
\begin{document}
\begin{center}
{\LARGE{\textbf{Role of Metal Nanoparticles on porosification of silicon by metal induced etching (MIE)}}}

\vspace{0.5 cm}
\textit{Shailendra K. Saxena$^{\$,a}$, Priyanka Yogi$^{\$,a}$, Pooja Yadav$^a$, Suryakant Mishra$^a$, Haardik Pandey$^a$, Hari Mohan Rai$^a$, Vivek Kumar$^b$, Pankaj R. Sagdeo$^a$ and Rajesh Kumar$^a$}\footnote{Corresponding author email: rajeshkumar@iiti.ac.in}

\vspace{0.5 cm}
$^a$ Material Research Laboratory, Discipline of Physics \& MSEG, Indian Institute of Technology Indore, Simrol-452020, Madhya Pradesh, India

$^b$ Department of Physics, National Institute of Technology Meghalaya, Laitumkhrah, Shillong-793003, Meghalaya, India

$^\$ $ \textit{Authors having equal contribution}

\vspace{1 cm}
ABSTRACT

\end{center}
Porosification of silicon (Si) by metal induced etching (MIE) process have been studies here to understand the etching mechanism. The etching mechanism has been discussed on the basis of electron transfer from Si to metal ion ($Ag^+$) and metal to $H_2O_2$. Role of silver nanoparticles (AgNPs) in the etching process has been investigated by studying the effect of AgNPs coverage on surface porosity. A quantitative analysis of SEM images, done using Image J, shows a direct correlation between AgNPs coverage and surface porosity after the porosification. Density of Si nanowires (NWs) also varies as a function of AgNPs fractional coverage which reasserts the fact that AgNPs governs the porosification process during MIE.
\vspace{0.5cm}


\section{Introduction}

Since its discovery, porous silicon(Psi)/Silicon nanostructures(SiNs) has become the topic of research due to its unique properties[1,2]. Amongst various methods used for fabrication of   Psi/SiNs [3-–6], metal induced chemical etching (MIE) is one of the simplest and cheapest methods for synthesis of semiconductor nanostructures[7–-9]. In the MIE method, firstly silver (Ag) or other metal nanoparticles (MNPs) are coated on a clean semiconductor surface usually by dipping the wafer in a solution containing appropriate metal salt dissolved in acid.  Size and distribution of AgNPs deposited on the wafer depends on concentration of the solution and reaction time. After the metal coating step, wafers (containing AgNPs) are transferred into etching solution for porosification. A typical etching solution for this purpose consists of hydrofluoric acid (HF) and hydrogen peroxide ($H_2O_2$). After a particular time of etching, Psi/SiNs get fabricated on the Si surface.

Different etching mechanisms have been proposed to understand the porosification process during MIE but unfortunately exact role of metal nanoparticles is not very clear.  Geyer at el[10] have given a model for the mass transport during MIE and Kolasinski [11] proposed an Electron transfer model on the basis of band bending at the metal/semiconductor interface and used different MNPs for etching of Si in MIE.  Backes et al[12] have proposed a model for MIE on the basis of the boundary condition of metal and semiconductor and discussed the mechanism by taking the various doping level of Si and different MNPs . Some researchers have shown the dual role of MNPs played in MIE [13]. They have discussed the role of metal ion and MNP distinctly. Smith et al[14] have discussed the mechanism of MIE on the basis of exchange current density between metal and semiconductor. Chartier et al[15] also discussed the mechanism of MIE on the basis of catalytic behavior of MNP. A clear demonstration is necessary for understanding the role of MNPs in the etching process during MIE.

In the present study, clear experimental evidence has been provided which suggests that MNPs govern the etching process. AgNPs have been used here on p-type Si wafer to fabricate our samples. The percentage covered surface area by AgNPs on the Si wafer is varied by the AgNPs deposition time from 15 s to 60 s. The obtained Si wafers with different coverage area (by AgNPs) were etched with a given etching time. SEM studies have been carried out to see the effect of AgNPs coverage on porosity of the Si wafers after MIE. Very good correlation between AgNPs coverage and surface porosity has been observed. These results and observations help to understand the mechanism of MIE
 
\section{Experimental Details}
P-type Si (111) wafers with resistivity of 0.01 $\Omega$-cm has been used to fabricate Psi/SNs using MIE method. Si wafer cut into four pieces and all the piece were cleaned in acetone and ethanol to remove impurities prior to starting the porosification process. The cleaned wafers were immersed in HF solution to remove any thin oxide layer formed at the Si surface. Then the wafers were dipped in solution containing 4.8 M HF \& 5 mM AgNO$_3$ for 15 s, 30 s, 45 s and 60 s at room temperature to deposit AgNPs. These four Ag NPs deposited samples were then kept for etching in an etching solution containing 4.8 M HF and 0.5 M $H_2O_2$ for etching time of 45 minutes. After the porosification process, the etched wafers were transferred in HNO$_3$ acid to remove extra Ag metal, After removal of the Ag metal, the samples were dipped into HF solution to remove any oxide layer induced by nitric acid used in the above step. List of all the samples under investigation has been provided in Table-1. Surface morphology of all the samples have been characterized using scanning electron microscopy (SEM) supra55 Zeiss and Carl Zeiss in both plan-view and cross-(X-) sectional geometries. Software Image J is used to calculate coverage area and surface porosity have been used to analyze the SEM results.
\begin{table}

 \caption{List of porous silicon samples with varying AgNPs deposition time}
\begin{center}
 \begin{tabular}{||c c c||} 
 \hline

 Sample Name & AgNP Deposition Time (s) &  Etching Time (Min) \\ [0.5ex] 
 \hline\hline
 P15 & 15 &  45 \\ 
 \hline
 P30 & 30 &  45 \\
 \hline
 P45 & 45  & 45 \\
 \hline
  P60 & 60 & 45  \\ [1ex] 
 \hline
\end{tabular}

\end{center}
\end{table}

\section{Results and Discussion}

Figure 1 shows the SEM images of Si wafers with AgNPs deposited with different deposition times in the range 15 seconds to 60 seconds. Surface morphologies of AgNPs covered Si wafer, shown in Fig. 1, reveals that increased AgNPs deposition time results in increased coverage area on Si wafer. As a result of increased AgNPs deposition time, the exposed Si wafer (uncovered area) surface decreases. The resultant exposed surface after this step decides the porosity and nanowire thickness as a result of MIE, which will be discussed later.  For the quantification of the deposition time dependent AgNPs coverage area, an estimation by the software ``Image J” has been carried out which shows that estimated percentages covered area increases from 41\% to 81 \% by increasing coating time from 15 seconds to 60 seconds respectively.

\begin{figure}
\begin{center}
\includegraphics[width=10cm]{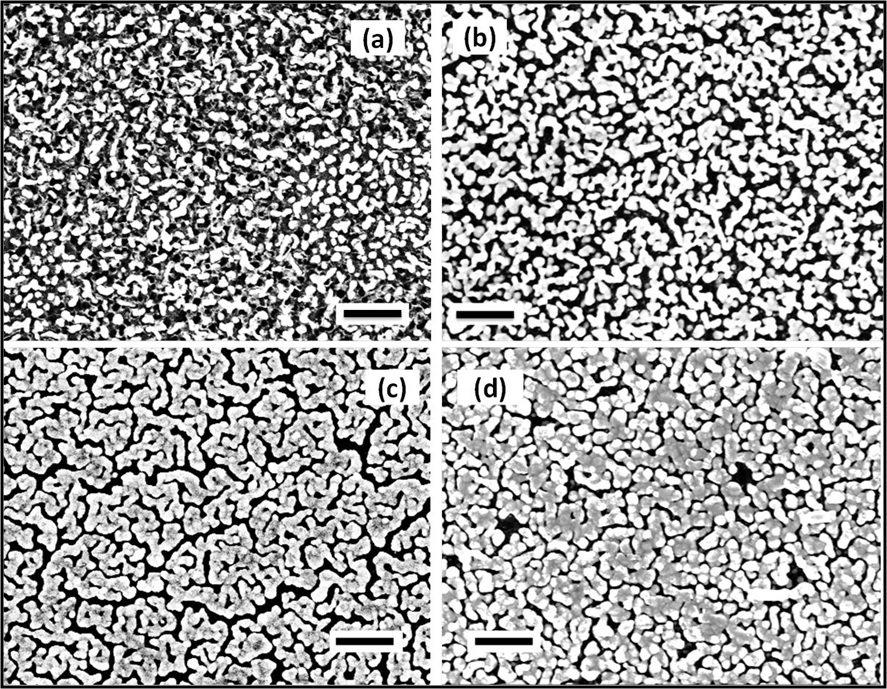}
\caption{SEM micrographs of silver coated Si surface for the coating time (a) 15 s (b) 30 s (c) 45 s (d) 60 s. Scale bars correspond to 500 nm.}
\end{center}
\end{figure}

To understand the effect of AgNPs on the process of porosification during MIE, above Si wafers, covered with AgNPs were immersed in etching solution for the formation of Psi/SiNs. Etching condition was the same for all the four samples with the etching time of 45 minutes. The SEM images showing surface morphologies of all the etched samples (as mentioned in Table 1) have been shown in Fig. 2. Porous surfaces have been obtained after MIE as can be seen in Fig. 2 which are similar to already reported by us [16–-18]. It is evident that pore diameters are not same for all the samples even though the etching time is constant. The pores look thinnest for sample P15 and widest for sample P60. It is also worth noting that sample P15 has been prepared from wafer where the AgNPs coverage was minimum and sample P60 has been prepared from a wafer where AgNPs coverage is maximum as discussed above. It indicates a direct correlation between the AgNPs coverage area and pore diameter and hence porosity.

\begin{figure}
\begin{center}
\includegraphics[width=10cm]{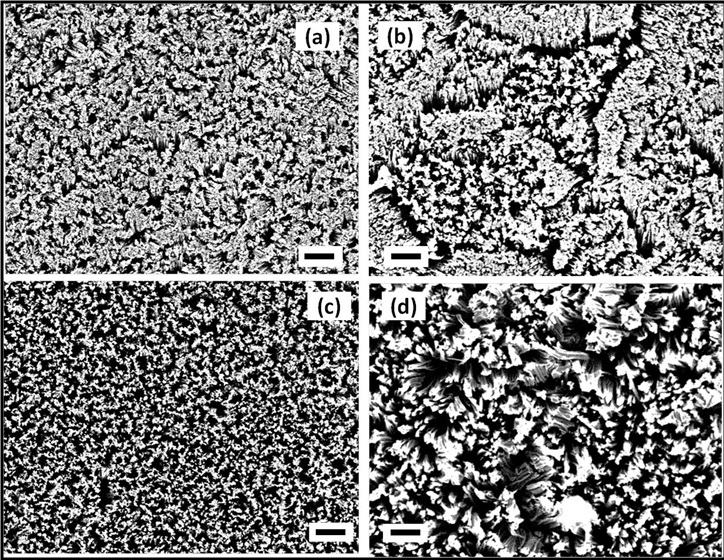}
\caption{Surface morphologies of etched Si surfaces for sample (a) P15 (b) P30 (c) P45 (d) P60. Scale bars correspond to 5 $\mu$m.}
\end{center}
\end{figure}

A quantitative analysis carried out using \textit{Inage J} which shows that the estimated surface porosities for all the sample increase from 35 \%  (for sample P15) to 74 \%  (for sample P60) as shown in Fig.3. Figure 3 also shows the variation in AgNPs coverage area, deposited prior to the porosification, as a function of AgNPs coating time. The trend in increasing surface porosity is in the same fashion with the increasing coverage area of Si by AgNPs which confirms the direct correlation between the surface porosity and fractional AgNPs coverage area.  The above-mentioned correlation, evident from Fig. 3, suggests that the etching mechanism for porosification involves the AgNPs as etching centers and dissolution of Si atoms starts from the sites of AgNPs which act as nucleation sites.   This can be elaborated on the basis of formation mechanism, as depicted in Fig.4 which is as follows. When clean Si wafers are dipped into the solution of AgNO$_3$ and HF available Ag$^+$ ions sit on the surface of Si wafer, take one electron from the Si surface and get neutralized into Ag. The transfer of electron from Si to Ag is possible because the redox energy for the pair Ag$^+$/Ag  (0.79 V) is higher than the valence band energy of Si (0.62 V) [19]. The Ag$^+$ accepts one electron from Si to create one hole (h$^+$) and gets deposited as  AgNPs on Si surface. It is understood that injection of hole inside the Si wafer is necessary for the porosification to start [8,20–-22].  The hole created during the above reaction initiates the following process when the wafer is kept in etching solution. The Ag deposited Si wafer when dipped into the etching solution (HF and $H_2O_2$) gets attacked by fluoride ion (F$^-$) at the positive site of Si created due to formation of hole. This forms $H_2SiF_6$ before getting dissolved in HF solution and creates a pit on the Si surface. Transfer of another electron from Ag to $H_2O_2$ results in the reduction of $H_2O_2$ and creation of Ag+ to start the cycle of etching by providing continuous supply of Ag$^+$ and thus of h$^+$. For this process, transfer of electron is favorable because the redox energy for the pair $H_2O_2 / H_2O$  (1.78 V) is higher than the redox energy for the pair Ag$^+$/Ag  (0.79 V)[19]. The etching time (continuation of cycle) decides the depth of pits in the Si). The etching time dependent study is already reported in the literature[22,23].  The mechanism discussed above is shown in the schematic diagram in Fig 4. It is worth mentioning here that the dissolution starts from the site where AgNPs are present because h+ are available there to initiate the etching process. A more AgNPs covered area will result in more porosification leading to highly porous surface. Above-mentioned etching mechanism clearly explains the observed correlation between AgNPs coverage area and surface porosity.

\begin{figure}
\begin{center}
\includegraphics[width=10cm]{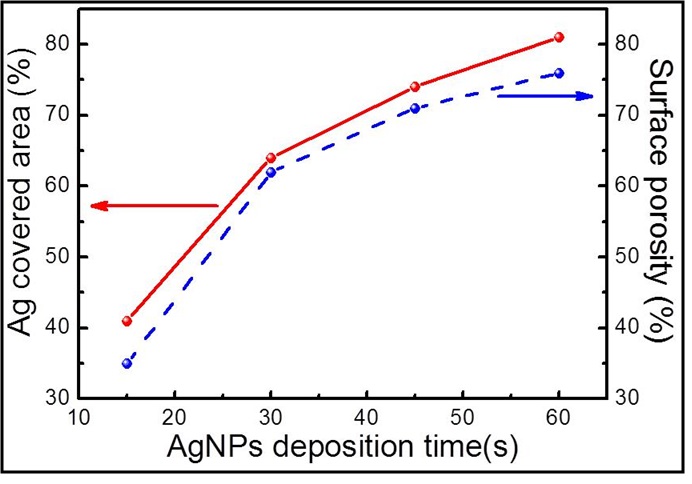}
\caption{The percentage of Ag covered area on Si and corresponding surface porosities.}
\end{center}
\end{figure}

\begin{figure}
\begin{center}
\includegraphics[width=10cm]{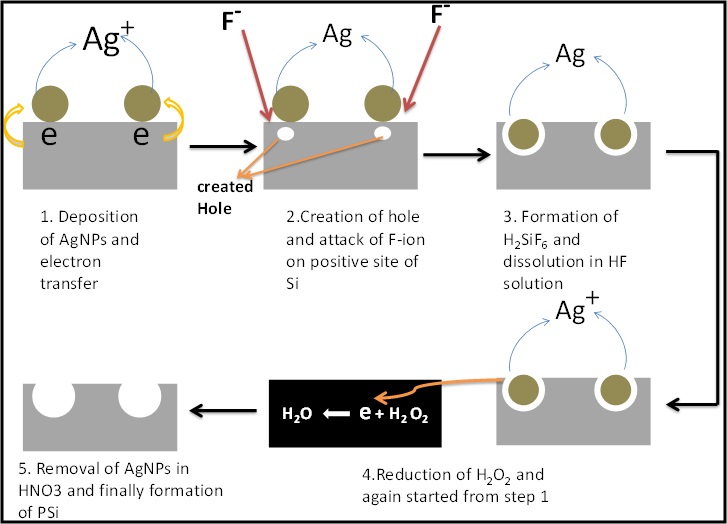}
\caption{Schematic diagram showing etching process during MIE of Si.}
\end{center}
\end{figure}

Progress of porisification in the vertical direction will also be interesting to analyze for understanding the etching process. The X-sectional SEM results for all the etched samples have been shown in Fig. 5. A comparative study of SEM images in figs 5(a) to (d) shows that large portion of Si wafer remains unetched for sample P15 as compared to other samples with higher AgNPs deposition times (samples P30, P45 and P60). Fig 5(a) also shows comparatively less density of vertical wire like structures (SiNWs). As the AgNPs coverage area increase more and more sites are available for initiation of etching and proceed in the vertical direction in the wafer. Si wafer with maximum coverage area was used to prepare Sample P60 where the surface porosity and pore density are maximum as can be revealed from Fig. 2(d) and Fig. 5(d).  As perceived from Fig 5, the quantity of unetched area decreases and the density of SiNWs increases on going from fig. 5(a) to Fig. (d).Abovementioned observation can be very clearly seen by comparing the X-sectional SEM images in Fig. 5 with their corresponding top view images in Fig.2. Effect of AgNPs coverage area on vertical etching can also be understood by taking the above mechanism (Fig. 4) in consideration. This also establishes the fact that the porosofication of Si by MIE is governed by AgNPs. 

\begin{figure}
\begin{center}
\includegraphics[width=10cm]{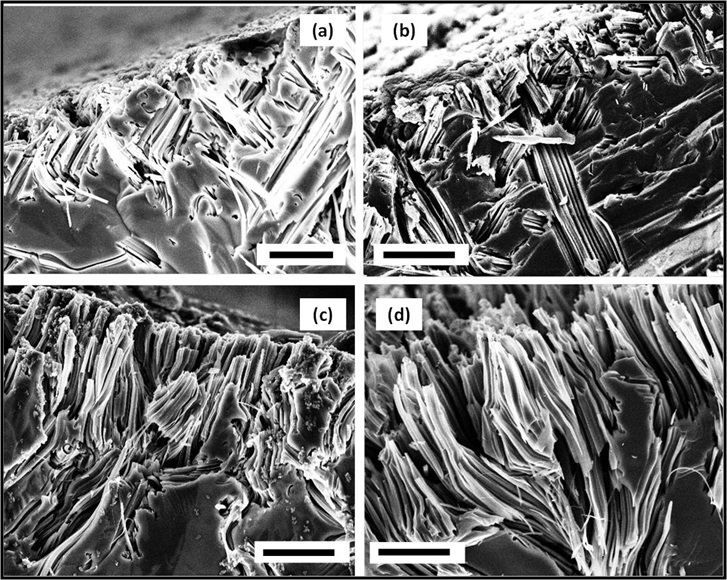}
\caption{Cross-sectional surface morphologies of etched Si surface for sample (a) P15 (b) P30 (c) P45 (d) P60. Scale bars correspond to 3 $\mu$ m.}
\end{center}
\end{figure}

\section{Conclusions}

In summary, various porous silicon samples prepared by changing AgNPs deposition time and a fixed etching time shows a direct correlation between the percentage surface porosity and AgNPs coverage area. The percentage of AgNPs covered area is estimated using Image J software and found to be in good agreement with the direct observation from the SEM results. Surface area coverage (by AgNPs) dependent surface porosity is observed. The role of MNPs in MIE has been discussed which implies that etching of the Si is mainly governed by MNPs. A variation in surface morphologies as a result of AgNPs coverage area reveals that etching process starts from the sites, where AgNPs are present. This is discussed in detail and an etching mechanism have been proposed for MIE. This model explains how not only the top morphology (surface porosity) but also the cross-sectional microstructure (SiNWs density) change during porosification using the  MIE process.  Coverage area dependent porosity variation of Si wafers reveals that AgNPs deposited prior to porosification governs the etching process during MIE.

\subsection*{Acknowledgement}
Authors acknowledge Sophisticated Instrumentation Center (SIC), IIT Indore for SEM measurements. Authors are also thankful to MHRD for providing fellowship. Authors acknowledge financial support from Department of Science and Technology (DST), Govt. of India.

\newpage


\begin{thebibliography}{31}
\bibitem{1}	A.G. Cullis, L.T. Canham, Nature 353 (1991) 335–338.
\bibitem{2}	O. Bisi, S. Ossicini, L. Pavesi, Surf. Sci. Rep. 38 (2000) 1–126.
\bibitem{3}	R. Kumar, H.S. Mavi, A.K. Shukla, Micron 39 (2008) 287–293.
\bibitem{4}	G. Sahu, D.P. Mahapatra, MRS Online Proc. Libr. 1354 (2011).
\bibitem{5}	L.T. Canham, Appl. Phys. Lett. 57 (1990) 1046–1048.
\bibitem{6}	K. Peng, Y. Yan, S. Gao, J. Zhu, Adv. Funct. Mater. 13 (2003) 127–132.
\bibitem{7}	Z. Huang, N. Geyer, P. Werner, J. de Boor, U. Gösele, Adv. Mater. 23 (2011) 285–308.
\bibitem{8}	X. Li, P.W. Bohn, Appl. Phys. Lett. 77 (2000) 2572–2574.
\bibitem{9}	L. Li, Y. Yao, Z. Lin, Y. Liu, C.P. Wong, in:, Electron. Compon. Technol. Conf. ECTC 2013 IEEE 63rd, 2013, pp. 581–585.
\bibitem{10}	N. Geyer, B. Fuhrmann, Z. Huang, J. de Boor, H.S. Leipner, P. Werner, J. Phys. Chem. C 116 (2012) 13446–13451.
\bibitem{11}	K.W. Kolasinski, Semicond. Sci. Technol. 31 (2016) 014002.
\bibitem{12}	A. Backes, A. Bittner, M. Leitgeb, U. Schmid, Scr. Mater. 114 (2016) 27–30.
\bibitem{13}	M. Abouda-Lachiheb, N. Nafie, M. Bouaicha, Nanoscale Res. Lett. 7 (2012) 455.
\bibitem{14}	Z.R. Smith, R.L. Smith, S.D. Collins, Electrochimica Acta 92 (2013) 139–147.
\bibitem{15}	C. Chartier, S. Bastide, C. Lévy-Clément, Electrochimica Acta 53 (2008) 5509–5516.
\bibitem{16}	S.K. Saxena, V. Kumar, H.M. Rai, G. Sahu, R. Late, K. Saxena, A.K. Shukla, P.R. Sagdeo, R. Kumar, Silicon (2015) 1–6.
\bibitem{17}	V. Kumar, S.K. Saxena, V. Kaushik, K. Saxena, A.K. Shukla, R. Kumar, RSC Adv. 4 (2014) 57799–57803.
\bibitem{18}	S.K. Saxena, R. Borah, V. Kumar, H.M. Rai, R. Late, V. g. Sathe, A. Kumar, P.R. Sagdeo, R. Kumar, J. Raman Spectrosc. (2015) n/a–n/a.
\bibitem{19}	Y. Qi, Z. Wang, M. Zhang, F. Yang, X. Wang, J. Phys. Chem. C 117 (2013) 25090–25096.
\bibitem{20}	H.S. Mavi, S. Prusty, M. Kumar, R. Kumar, A.K. Shukla, S. Rath, Phys. Status Solidi -Appl. Mater. Sci. 203 (2006) 2444–2450.
\bibitem{21}	C.A. Betty, R. Lal, D.K. Sharma, J.V. Yakhmi, J.P. Mittal, Sens. Actuators B Chem. 97 (2004) 334–343.
\bibitem{22}	L. Lin, S. Guo, X. Sun, J. Feng, Y. Wang, Nanoscale Res. Lett. 5 (2010) 1822.
\bibitem{23}	S.K. Saxena, G. Sahu, V. Kumar, P.K. Sahoo, P.R. Sagdeo, R. Kumar, Mater. Res. Express 2 (2015) 036501.

\end{thebibliography}
\end{document}